\documentclass{mn2e}
\usepackage{graphicx}
\input epsf

\newcommand{\Fs}{\,^*\! F}

\newcommand{\bJ}{\bmath{J}}

\newcommand{\bB}{\bmath{B}}
\newcommand{\bE}{\bmath{E}}
\newcommand{\bH}{\bmath{H}}

\newcommand{\text}[1]{\quad\mbox{#1}\quad}

\newcommand{\vpr}[2]{\bmath{#1} \!\times\! \bmath{#2}}

\newcommand{\vcurl}[1]{\vpr{\nabla}{#1}}

\newcommand{\Pd}[1]{\partial_{#1}}

\newif\ifAMStwofonts

\title
{Observations of the Blandford-Znajek and the MHD Penrose
processes in computer simulations of black hole magnetospheres}  
\author[S.S. Komissarov]
{
  S.S.Komissarov\\
Department of Applied Mathematics, 
The University of Leeds, 
Leeds LS2 9JT }

\begin{document}
\label{firstpage}
\maketitle

\begin{abstract} 

{In this paper we report the results of axisymmetric relativistic 
MHD simulations for the problem of Kerr black hole immersed into a 
rarefied plasma with ''uniform'' magnetic field. The long term solution
shows properties which are significantly different from those of 
the initial transient phase studied recently by Koide(2003). 
The topology of magnetic field lines within the ergosphere is similar to 
that of the split-monopole model with a strong current sheet in the equatorial 
plane. Closer inspection reveals a system of isolated magnetic islands inside 
the sheet and ongoing magnetic reconnection. 
No regions of negative hydrodynamic ''energy at infinity'' are seen 
inside the ergosphere and the so-called MHD Penrose process does not 
operate. Yet, the rotational energy of the black hole continues to be 
extracted via purely electromagnetic mechanism of Blandford and Znajek(1977). 
However, this is not followed by development of strong relativistic outflows 
from the black hole. Combined with results of other recent simulations
this signals a potential problem for the standard MHD model of relativistic 
astrophysical jets should they still be observed at distances as small as 
few tens of gravitational radii from the central black hole.    }
\end{abstract}


\section{Introduction}

Observations of various astrophysical phenomena such as 
active galactic nuclei and galactic microquasars often reveal
powerful relativistic jets streaming away from a massive central object, 
most likely a black hole. 
The exact mechanism of generating powerful relativistic jets from 
black holes system is not yet known, although a number of interesting 
models have been put forward and 
studied with various degree of detail during the last few decades.     

By now there has emerged a general consensus that the central engine 
should involve a rotating black hole linked with the jets by means 
of strong magnetic field and an accretion disc supporting this magnetic
field by its electric currents. This magnetic field is believed to serve a number 
of key functions: 1) to power the jets via extracting the rotational energy 
of the black hole; 2) to provide the jet collimation via magnetic hoop stress; 
and 3) to suppress mixing of the rarefied jet plasma with the relatively dense 
surrounding medium such as the coronal plasma of the accretion disc, which is 
needed to ensure that the jet plasma remains magnetically dominated and, thus, 
can be accelerated to the ultra-relativistic speeds inferred from the 
observations.  

The most simplified mathematical frameworks for highly magnetised 
relativistic plasma are an 
ideal relativistic magnetohydrodynamics (MHD) and magnetodynamics 
(MD, which can be described as MHD in the limit of zero particle 
inertia, e.g Komissarov 2002). Both frameworks, however, involve rather 
complex equations particularly in the case of curved spacetime. 
This explains a relatively slow progress in the 
theory -- only a limited number of analytical solutions have been found 
so far and only for the limiting case of a slowly rotating black hole where 
one can employ the perturbation method. The most important of them 
is the MD solution for a monopole magnetosphere by Blandford and 
Znajek (1977). Indeed, this solution describes an  outgoing Poynting flux 
from a Kerr black hole as if the hole was a magnetized rotating conductor whose 
rotational energy can be efficiently extracted by means of magnetic torque. 
This most important feature of the Blandford-Znajek solution (BZ) is also the
most intriguing and puzzling as such conductor does not really exist. 

Often, this missing element of the BZ model is artificially introduced in the form of
the so-called ``membrane'', or the ``stretched horizon'', located
somewhat above the real event horizon \cite{MT82,TPM}. 
This ''physical'' interpretation of the BZ solution provides 
simple means of communicating the results to wide astrophysical community and 
for this reason it has been almost universally accepted. However, its artificial 
nature and the fact that no proper physical interpretation had been pushed forward 
also made the result vulnerable to criticism on theoretical grounds and 
stimulated attempts to find alternative ways of magnetic extraction of rotational energy 
of black holes \cite{Pun-Cor,Pun01,Tak}. These attempts seem to draw the
inspiration from a completely non-magnetic mechanism proposed much earlier 
by Penrose (1969). The key role in this mechanism is played by the ergosphere 
of a rotating black hole. Within the ergosphere particles can acquire negative 
energy (or rather ''energy at infinity'') and if such particles are swallowed
by the black hole its energy can decrease. 
In the original Penrose process such particles are created via 
close range interaction 
(collisions, decay) with other particles which gain positive energy and carry 
it away. 
However, electrically charged particles can also be pushed onto orbits with 
negative energy by the Lorentz force and this is what makes possible the so-called 
''MHD Penrose process'' \cite{Tak,Koi02,Koi03} which is the common element of 
all alternative magnetic mechanisms.

Recently there has been a renewed interest to this problem. The main reason for 
this is the arrival of robust numerical methods for relativistic astrophysics, e.g.
\cite{Pons,Kom99,Koi99,Kol,Gam,Del-Zan,Dev-Haw03}. Time-dependent MD and MHD 
simulations of monopole magnetospheres of black holes have demonstrated the  
asymptotic stability of the BZ solution \cite{Kom01} and provided additional 
arguments in favour of the MD approximation in this particular case \cite{Kom04a,Kom04b}. 
Moreover, simulations of magnetically driven accretion disks have shown the 
development of low density axial regions, ''funnels'', closely 
described by the BZ solution \cite{Dev03,McK}.  

On the other hand, the MHD simulations of a black hole immersed into a rarefied 
plasma with uniform magnetic field seemed to provide support for the 
MHD Penrose model \cite{Koi02,Koi03}.
At least, it was found that by the end of the simulations the inflow of 
particles with negative energy at infinity accounted for about one half 
of the extracted energy. Unfortunately, the termination time of these simulations 
was surprisingly short, approximately one half of the rotational period of 
the black hole (presumably due to some computational problems.)  
As the result one could not tell whether the MHD Penrose mechanism would remain 
effective on a long-term basis. In this paper we present the 
results of new simulations which were carried out for much longer period of time 
and which bring new light on this important issue.

\section{Basic equations and numerical method}

In these simulations we solve numerically the equations of ideal MHD in the 
space-time of a Kerr black hole. This space-time is described using the foliation 
approach where the time coordinate $t$ parametrises a suitable filling 
of spacetime with space-like hypersurfaces described by the 3-dimensional 
metric tensor $\gamma_{ij}$. These hypersurfaces may be regarded as the
``absolute space'' at different instances of time $t$. 
If $\{x^i\}$ are the spatial coordinates of the absolute space then the 
metric form can be written as 
\begin{equation}
  ds^2 = (\beta^2-\alpha^2) dt^2 + 2 \beta_i dx^i dt +
         \gamma_{ij}dx^i dx^j
\label{metric}
\end{equation}
where $\alpha>0$ is called the ``lapse function'' and
$\bbeta$ is the ``shift vector''. In this study we employ the Kerr-Schild
coordinates, ${t,\phi,r,\theta}$, which like the well known Boyer-Lindquist 
coordinates ensure that none of the components of the metric form depend on 
$t$ and $\phi$. At spatial infinity these coordinate systems do not 
differ but the Kerr-Schild system does not have a coordinate singularity 
on the event horizon. Other details can be found in e.g. 
\cite{Kom04a,McK}. 

The evolution equations of ideal MHD include  the continuity equation,

\begin{equation}
\Pd{t}(\alpha\sqrt{\gamma}\rho u^t)+ \Pd{i}(\alpha\sqrt{\gamma}\rho u^i)=0,
\label{cont1}
\end{equation}
the energy-momentum equations,
                                                                                          
\begin{equation}
\Pd{t}(\alpha\sqrt{\gamma}T^t_{\ \nu})+ \Pd{i}(\alpha\sqrt{\gamma}T^i_{\ \nu})=
\frac{1}{2} \Pd{\nu}(g_{\alpha\beta}) T^{\alpha\beta} \alpha\sqrt{\gamma},
\label{en-mom1}
\end{equation}
and the induction equation,
                                                                                          
\begin{equation}
\Pd{t}(B^i)+e^{ijk}\Pd{j}(E_k) =0.
\label{ind1}
\end{equation}
Here $g_{\alpha\beta}$ is the metric tensor of spacetime,
$\gamma=\det(\gamma_{ij})$, \\ $e^{ijk}$ is the Levi-Civita
pseudo-tensor of space, $\rho$ is the proper mass density of plasma,
$u^\nu$ is its four-velocity vector.  The total stress-energy-momentum
tensor, $T^{\mu\nu}$, is a sum of the stress-energy momentum tensor of
matter,
\begin{equation}
   T_{(m)}^{\mu\nu} = wu^\mu u^\nu -p g^{\mu\nu},
\end{equation}
where $p$ is the thermodynamic pressure and $w$ is the enthalpy per
unit volume, and the stress-energy momentum tensor of electromagnetic
field,
\begin{equation}
   T_{(e)}^{\mu\nu} = F^{\mu\gamma} F^\nu_{\ \gamma} -
   \frac{1}{4}(F^{\alpha\beta}F_{\alpha\beta})g^{\mu\nu},
\end{equation}
where $F^{\nu\mu}$ is the Maxwell tensor of the electromagnetic field.
The electric field, $\bE$, and the magnetic field, $\bB$, are defined
via
\begin{equation}
  E_i=\frac{\alpha}{2} e_{ijk}\Fs^{jk} ,
\end{equation}
and
\begin{equation}
  B^i=\alpha \Fs^{it},
\end{equation}
where $\Fs^{\mu\nu}$ is the Faraday tensor of the electromagnetic
field, which is simply dual to the Maxwell tensor. In the limit of
ideal MHD

\begin{equation}
  \bE=-\vpr{v}{B} \text{or} E_i=e_{ijk}v^jB^k,
\label{perf-cond}
\end{equation}
where $v^i=u^i/u^t$ is the usual 3-velocity of plasma. 
In addition, the magnetic field must satisfy the divergence free condition

\begin{equation}
  \Pd{i}(\sqrt{\gamma} B^i) =0 .
\end{equation}
Note, that 1) all the
components of vectors and tensors appearing in these equations are the 
those measured in the coordinate basis, $\{\Pd{\nu}\}$, of the Kerr-Schild 
coordinates; 2) throughout the paper we employ such units that the speed 
of light $c=1$, the gravitational constant $G=1$, the black hole mass $M=1$, 
and the factor $4\pi$ does not appear in Maxwell's equations.

Our numerical scheme is a 2D Godunov-type upwind scheme which utilises  
special relativistic Riemann solver described in Komissarov(1999) and   
uses the method of constraint transport \cite{Eva-Haw} to preserve the 
magnetic field divergence free. Other details are outlined in 
\cite{Kom04b}. 

\begin{figure}
\includegraphics[width=85mm]{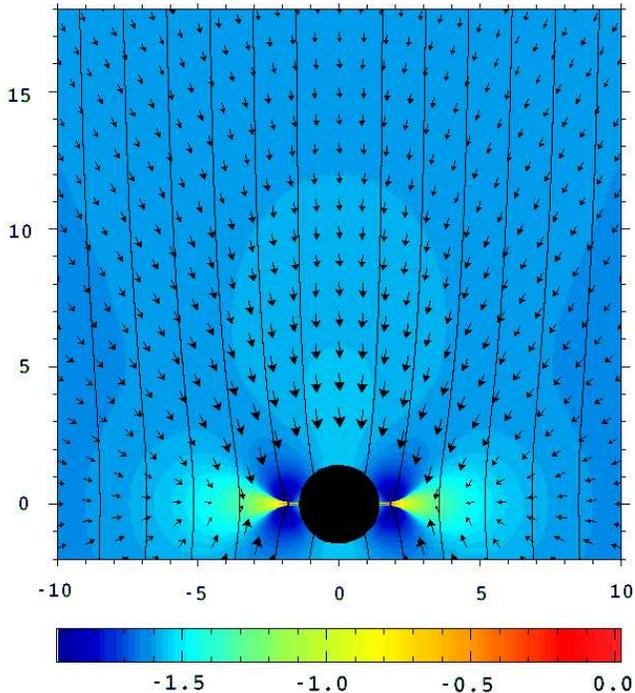}
\caption{The large scale structure of the numerical solution during the initial 
phase, $t=6$. The colour image shows the distribution of rest mass density 
,$ \log_{10} \rho$. The contours show the magnetic flux surfaces (the lines of the 
poloidal component of magnetic field.) The arrows show the poloidal component 
of velocity relative to the coordinate grid.}
\label{init}
\end{figure}

\section{Numerical simulations}

\subsection{Setup}

In these simulations the rotational parameter $a$ of the Kerr metric is 
$a=0.9$ which gives the event horizon radius $r_+ \simeq 1.44$.    
The axisymmetric computational domain covers $ 0<\theta<\pi$ and 
$1.35<r<53.1$. The computational grid has 401 cells in the $\theta$-direction, 
where it is uniform, $\Delta\theta = \mbox{const}$ , and 400 cells in the $r$-direction. 
The cell size in the $r$-direction, $\Delta r$, is such that the
corresponding physical lengths in both directions are equal in the equatorial plane.  

The usual axisymmetric boundary conditions are imposed at $\theta=0$ and 
$\theta=\pi$ boundaries. At the outer boundary, $r=53.1$, the initial values 
of all variables are imposed throughout the whole run, the termination time 
for the simulations was set to $t=60$ so no waves emitted from the dynamically
active region near the black hole had a chance to get reflected of the 
boundary and effect the inner solution. 
The inner boundary $r=1.35$ is well inside the event horizon, 
which justifies use of ``radiative boundary conditions''.

The initial velocity field is set to be the same as the one of the 
fiducial observers of the Kerr-Schild foliation, who spiral towards 
the black hole (e.g. Komissarov 2004a).

The initial electromagnetic field has the same ``uniform'' magnetic 
component, aligned with the rotational axis of the black hole, as in the 
vacuum  solution of Wald: 
\begin{equation}
   F_{\mu\nu}= (m_{[\mu,\nu]} +2a k_{[\mu,\nu]}),
\label{wald}
\end{equation}
where $ k^\nu = \Pd{t}$ and $m^\nu = \Pd{\phi}$ are the Killing vectors
of the Kerr spacetime \cite{Wald}. However, the initial electric component 
is different as it has to satisfy the condition of perfect conductivity 
(\ref{perf-cond}). 

 The initial thermodynamic pressure and the rest mass 
density of plasma are set to be $p=0.01 p_m$ and $\rho=0.05 p_m$, where 
$p_m=(B^2-E^2)/2$ is the magnetic pressure. Finally, the 
equation of state employed in the simulations describes polytropic gas 
with the ratio of specific heats, $\Gamma=4/3$.

Summarising,  although the setup of our simulations is not exactly 
the same as in  Koide\shortcite{Koi03} it is quite similar. In both 
cases we are dealing with magnetically dominated plasma of similar 
magnetisation. In both cases the initial magnetic field is described 
by the Wald vacuum solution \cite{Wald}. Thus one would expect to obtain 
at least qualitatively 
similar solution in the common region of computational domains. Because of 
different space-time foliations, Koide\shortcite{Koi03} used the Boyer-Lindquist 
coordinates, 
the difference in computational domains is not simply reduced to the difference 
in the range of the radial coordinate, $r$. Different definitions of the global time
coordinate, $t$, should also be taken into account. This, however, is only significant
in the spacial regions very close to the event horizon where the Boyer-Lindquist
system becomes singular.    

\begin{figure*}
\includegraphics[width=178mm]{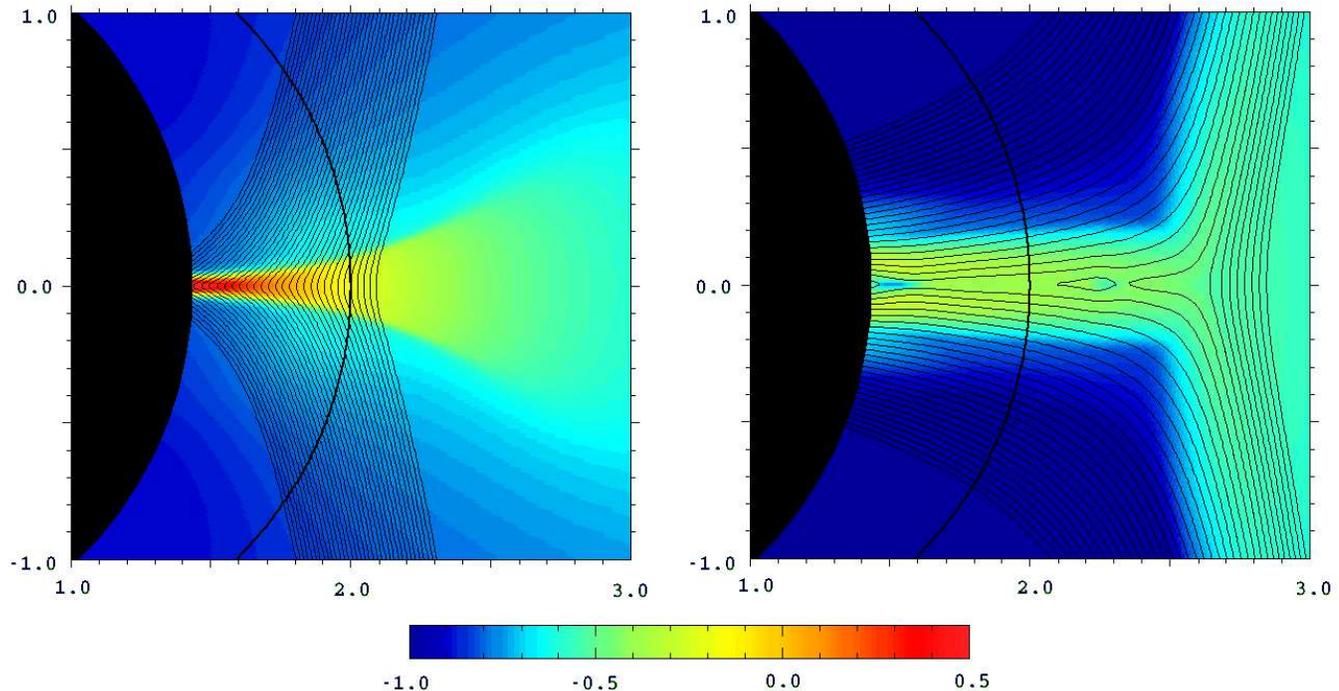}
\caption{Left panel: The $u_t$ component of plasma four-velocity and magnetic 
flux surfaces at $t=6$. Right panel: the same as in the left panel but at $t=60$. 
In both panels there is shown the same set of magnetic flux surfaces.   
The thick solid line shows the ergosphere.
Note that $-u_t$ gives the specific ''energy at infinity'' for 
a free particle and that the region of negative hydrodynamic energy at infinity 
is even somewhat narrower than the region of positive $u_t$ because of the 
pressure contribution. }
\label{comp}
\end{figure*}

All conservative schemes for the relativistic magnetohydrodynamics have 
an upper limit on plasma magnetization above which they fail. At
this limit, which somewhat varies from problem to problem and also depends 
on the resolution, the numerical error for
the total energy density becomes comparable with the energy density of 
matter.  This forces us to pump in fresh plasma in regions where 
its magnetization becomes dangerously high. Since in such regions the 
dynamical role of particles is rather insignificant this measure should 
not have a strong effect in most respects. The critical condition we 
set in these simulations is
\begin{equation} 
   wW^2 - p = 0.01 B^2,
\label{cond} 
\end{equation}
where $W$ is the Lorentz factor of the flow and $B$ is the magnetic 
field strength as measured by the local FIDO. 
Should the energy density of matter drop below $0.01 B^2$, both $\rho$ and
$p$ are artificially increased by the same factor.  To minimise the
effect of the mass injection on the flow the velocity of
the injected matter is set to be equal to the local velocity of the flow.
In fact, new particles must be constantly created in real
magnetospheres of black holes but the details of this process can be
rather different \cite{Bes,Hir-Oka,Phi}.
An additional lower limit was set on the value of the thermodynamic
pressure, which was not allowed to drop below $0.01\rho$.

\begin{figure}
\includegraphics[width=85mm]{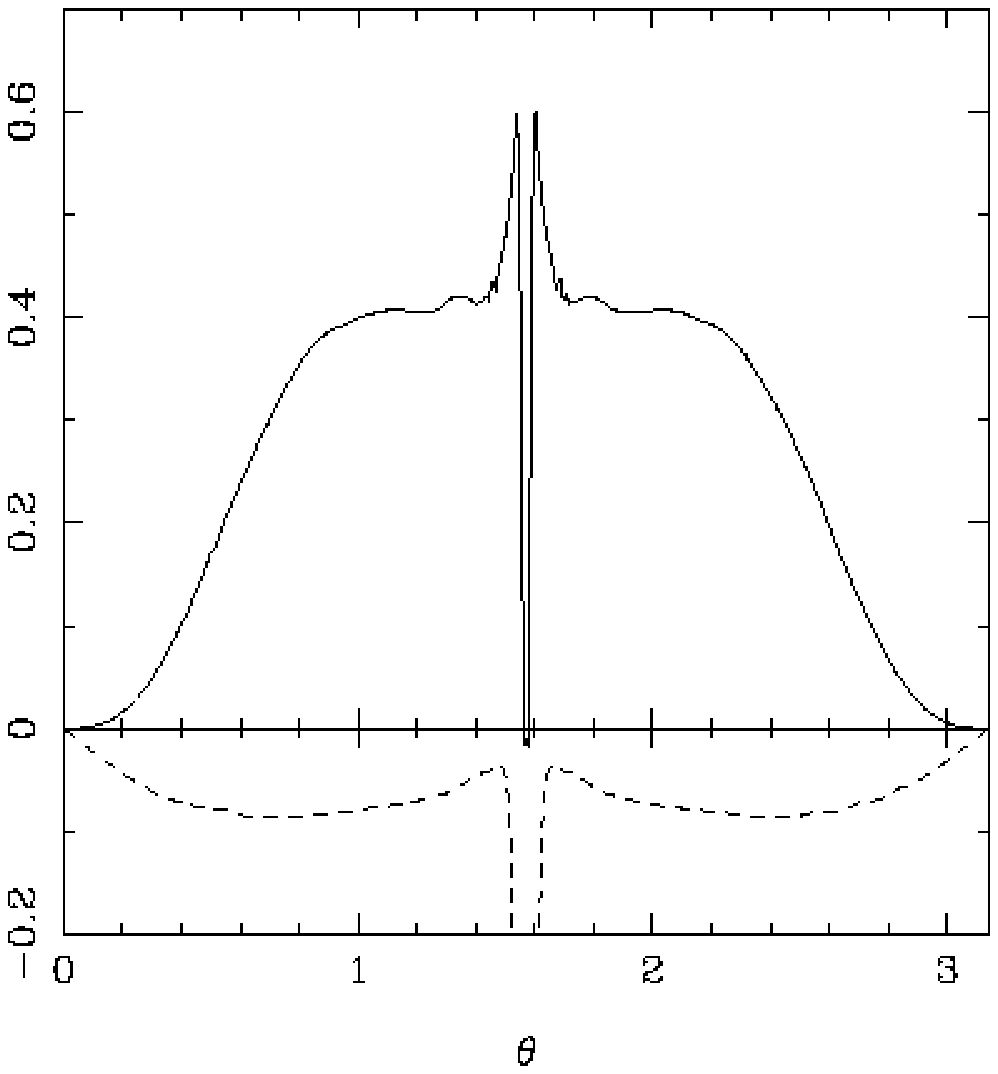}
\caption{The hydrodynamic (dashed line) and the electromagnetic 
(solid line) components of the energy flux ($-T^r_{\ t}$) through the event 
horizon  at $t=60$.}
\label{NoPenrose}
\end{figure}

\subsection{Results and Discussion}

We have found that numerical solution exhibits two phases with rather 
different properties: 1) the rather short initial phase which is dominated  
by a rapid evolution in the neighbourhood of the black hole and 2) the final
phase where solution settles to an approximate steady state in this region.    

\begin{figure*}
\includegraphics[width=85mm]{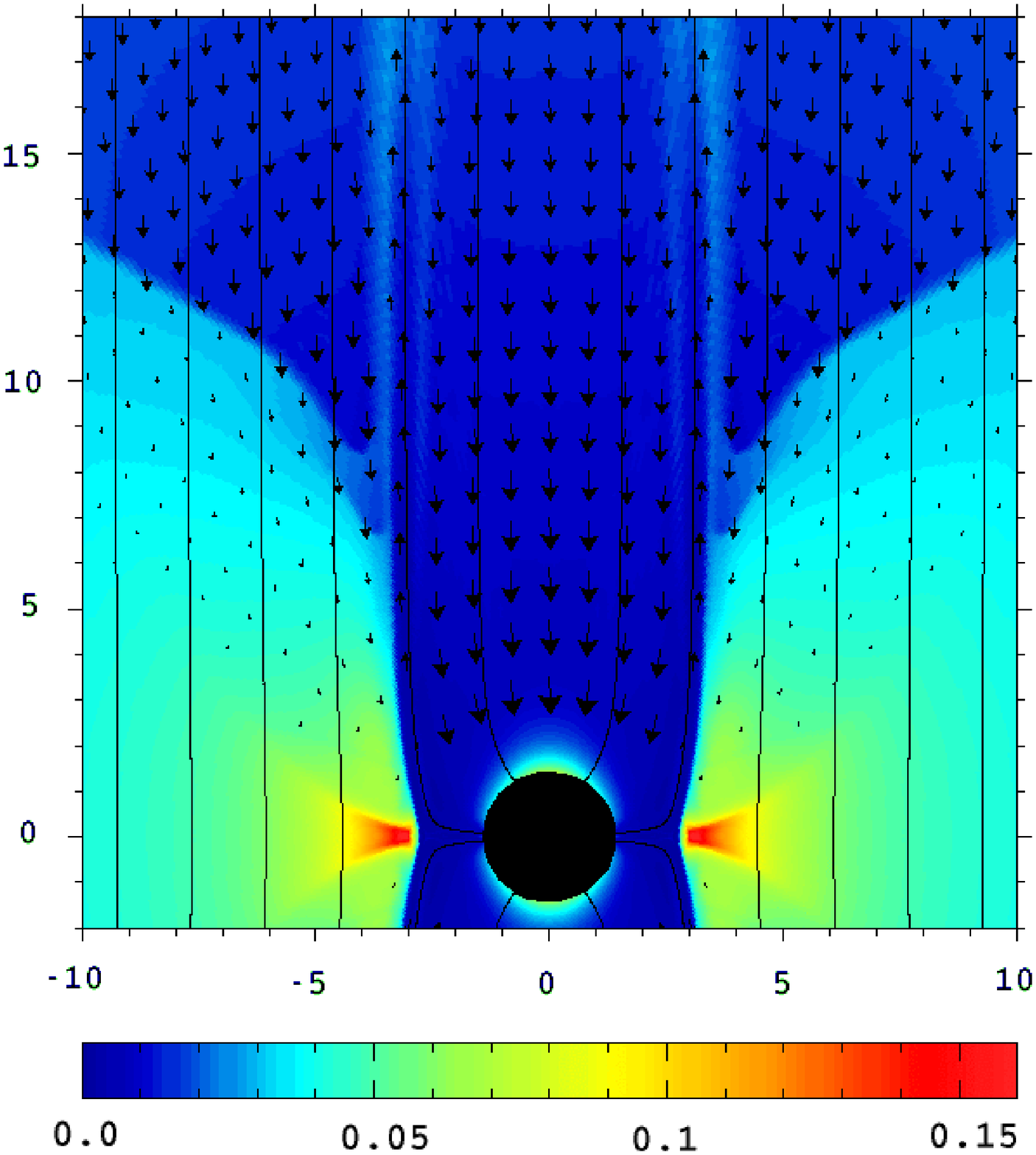}
\includegraphics[width=85mm]{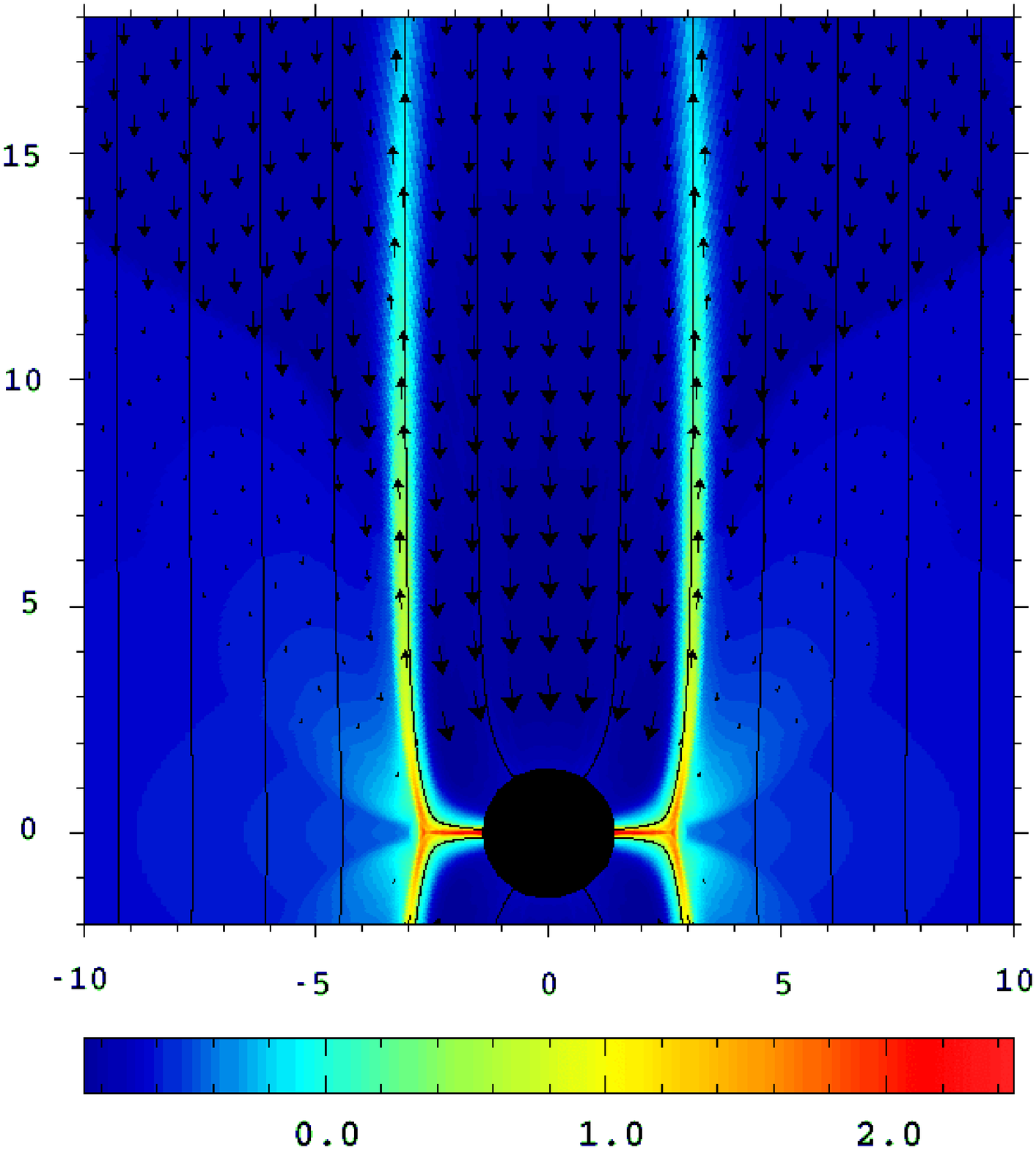}
\includegraphics[width=85mm]{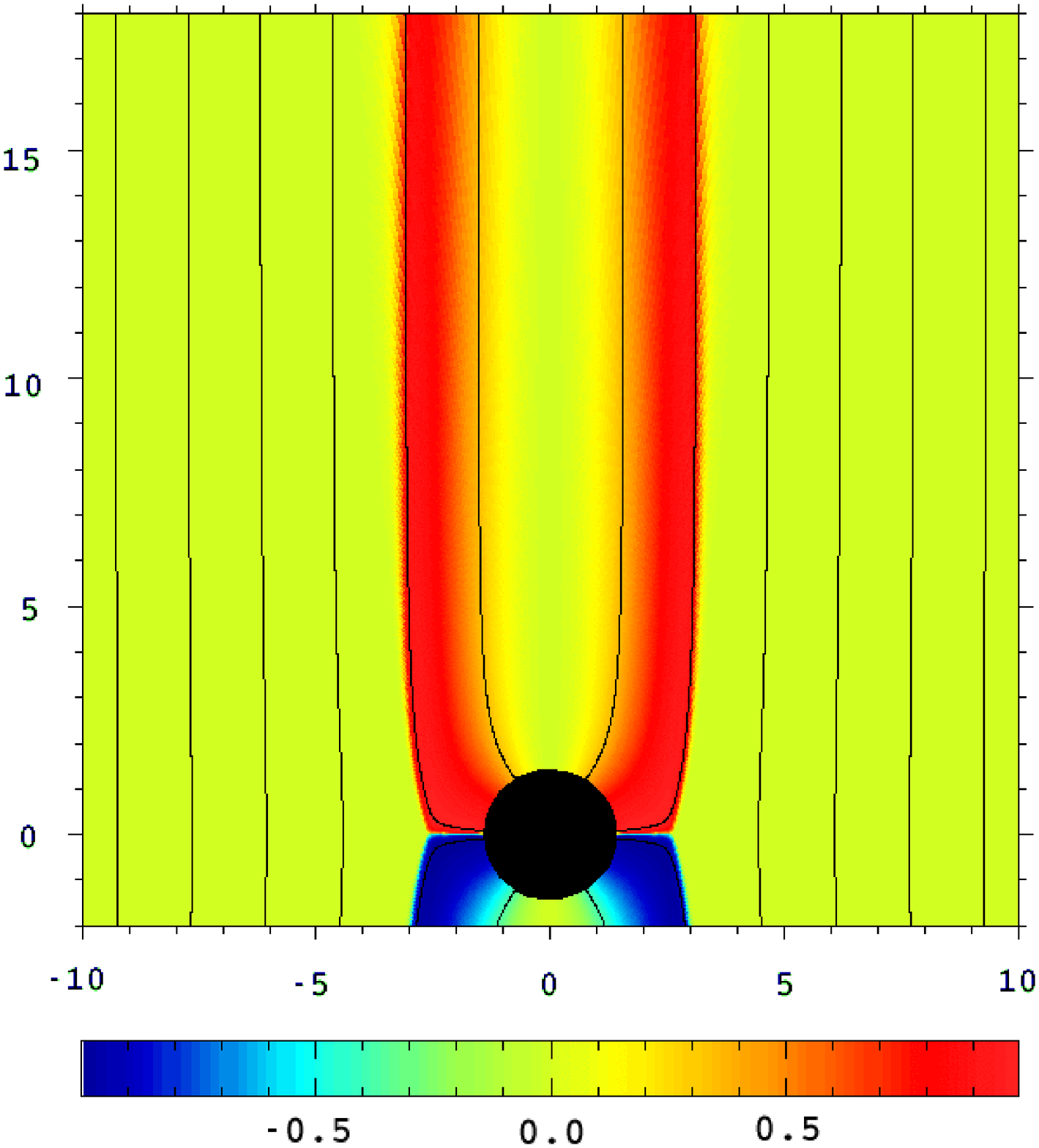}
\includegraphics[width=85mm]{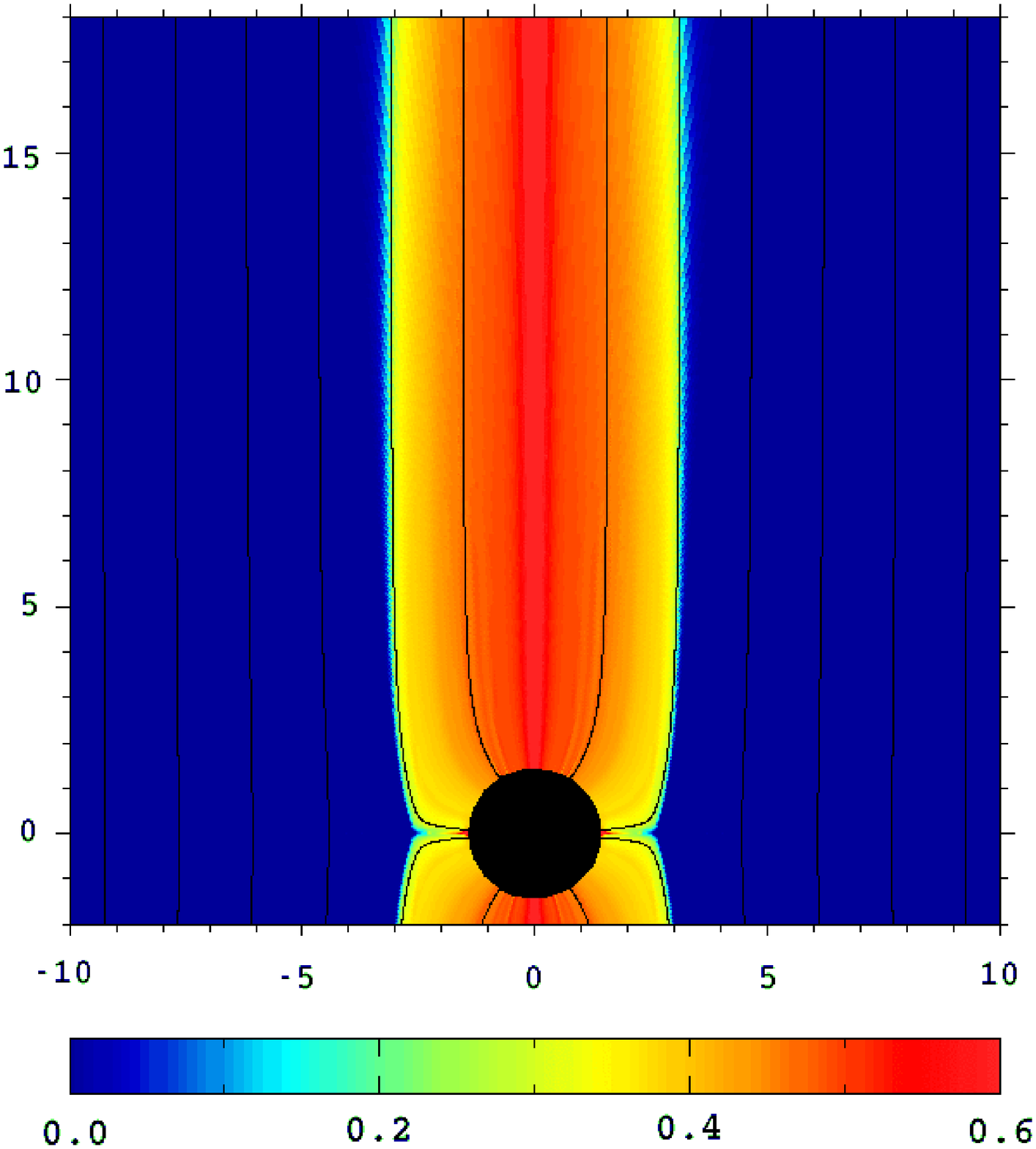}
\caption{The large-scale structure of the solution at $t=60$. 
{\it Top left panel:} the rest mass density, $\rho$; 
{\it Top right panel:} the gas temperature, $\log_{10} (P/\rho)$; 
{\it Bottom left panel:} $H_\phi$=(the total electric current through the 
circular loop $r=$const)$/2\pi$; 
{\it Bottom left panel:} the angular velocity of magnetic field lines 
$\Omega/\Omega_h$, where $\Omega_h$ is the angular velocity of the 
black hole. The solid lines show the magnetic flux surfaces and the 
arrows show the poloidal velocity relative to the coordinate grid of
the Kerr-Schild coordinates.   
}
\label{final}
\end{figure*}

In the initial phase our numerical solution is indeed very similar to
the one described in Koide \shortcite{Koi03}. Plasma mainly slides 
along the magnetic field lines towards the equatorial plane where it 
passes through an accretion shock and forms an equatorial disc 
(see fig.\ref{init}). The magnetic field lines are also pulled towards 
the black hole. Further away from the hole this seems to be mainly due the 
velocity field of the initial solution as the subsequent increase of the magnetic
pressure in the central region quickly halts this motion and the magnetic 
field lines of these distant regions eventually straighten up 
(see fig.\ref{final}).  However, near the horizon  
this pulling is certainly caused by the hole as the field lines 
never straighten up there. Since similar effect is also observed in 
simulations of inertia-free  magnetospheres \cite{Kom04a} 
it cannot not be explained simply by dragging of the magnetic field lines 
along with the accreted plasma and has to have a more general cause.  
In any case, these findings are in strong contrast with the exclusion of 
magnetic flux by rotating black holes found in axisymmetric vacuum solutions 
\cite{Wald,BJ}. Those vacuum solutions were used  in the past to argue low 
efficiency of the BZ mechanism in the case of rapidly rotating black holes. 
Our results shows that this argument is incorrect and vacuum solutions have to 
be used with more caution.

The left panel of figure \ref{comp} shows the inner region of our solution 
at $t=6$ in more detail. The colour image describes the distribution of 
the covariant time component of plasma's four-velocity, $u_t$, whereas 
the solid lines show the magnetic flux surfaces (the lines of poloidal magnetic 
field). For a free particle $-u_t$ gives its specific energy at infinity.  Since 
the hydrodynamic part of energy at infinity has the volume density 

\begin{equation}
   -\alpha T^t_{\ t_{(m)}} = \alpha (-w u_t u^t + p)
\end{equation}
it can only be negative if $u_t>0$; because of the pressure 
contribution  the region of negative hydrodynamic energy at infinity is 
somewhat narrower than the region of positive $u_t$.  What is even more important 
is that the flux of hydrodynamic energy at infinity

\begin{equation}
    -\alpha T^i_{\ t_{(m)}} =-\alpha w u_t u^i
\end{equation}
has the opposite direction to the flow velocity $v^i = u^i/u^t$, as 
required in the MHD Penrose process, if and only if $u_t$ is positive. 
Figure~\ref{comp} shows that well in agreement with findings of \cite{Koi02,Koi03} 
the inner part of the ergospheric disc does indeed have  positive $u_t$ during 
the initial phase.  
Moreover, the magnetic field has a very similar topology too -- all magnetic
field lines threading the ergospheric disc have a turning point in the equatorial 
plane and do not cross the event horizon.

The reason for developing positive $u_t$ looks attractively simple. Due to the inertial 
frames dragging, all plasma entering the black hole ergosphere is forced to rotate in the 
same sense as the black hole (to be more precise this is what is observed by a  
distant observer. In Kerr-Schild coordinates $d\phi/dt$ may have both signs.) 
As the result a differential rotation inevitably develops along the magnetic 
field lines penetrating the ergosphere which causes twisting of these lines. 
The Alfven waves generated in this manner  propagate away from the ergosphere 
to infinity and establish an outflow of energy in the form of 
Poynting flux. Because of the energy conservation the ergospheric plasma 
reacts by moving onto orbits of lower energy.  Provided this plasma remains in the space between the event horizon and the ergosphere for long enough 
it may indeed end up having negative energy.

In the problem under consideration the strong magnetic field keeps the plasma 
of the ergospheric disc from falling into the black hole and allows it 
to gain negative energy (see fig.\ref{comp}). 
However, such configuration cannot be sustain forever within ideal MHD. Since, 
the energy is constantly extracted along the field lines of the equatorial 
disc the energy of the disc is constantly going down and no steady-state
can be reached. At some point the magnetic configuration would have to 
change so that all magnetic field lines entering the ergosphere also penetrate 
the event horizon (Another possible option could be a non-steady behaviour with 
magnetic field lines pulled in and out of the event horizon all the time.) 
As one can see in the right panel of figure~\ref{comp}, where we present the 
solution at $t=60$, this is more or less what is observed in our simulations. 
At around $t=20$ the ergospheric disc is fully swallowed by the hole and a strong 
current sheet develops in its place. As the result, the structure of magnetic 
field becomes similar to that of the split-monopole model \cite{BZ}.  
This configuration persists till the very end of the simulations when the 
whole solution in the inner part of the spacial domain seems to reach a state 
of an approximate equilibrium.

In fact, one can see in fig.\ref{comp} at least two magnetic islands in the 
equatorial current 
sheet. Such structure is known be quite typical for the tearing mode instability 
(e.g. Priest \&  Forbes, 2000) and was suggested in the context of black hole 
magnetospheres in Beskin (2003).  
Here, the islands are formed during slow reconnection events resulting in 
the gradual escape of magnetic flux from the event horizon, the inner islands 
disappear into the hole whereas the outer ones move away from it 
and supply hot plasma for a thin outflow sheath clearly seen in 
the temperature plot in fig.\ref{final}. This evolution is indeed very slow and 
might be related to a somewhat  excessive capture of magnetic field line during 
the early stages.  

The most important feature of the solution after the restructuring 
of magnetic field is the total disappearance of the regions with 
positive $u_t$ (see the right panel of fig.\ref{comp}). Thus, the MHD 
Penrose process does no longer operate in the black hole ergosphere.  
This is confirmed in figure~\ref{NoPenrose} which shows that the hydrodynamic 
flux of energy at infinity through the event horizon,  
$-\alpha T^r_{\ t_{(m)}}$, is everywhere negative. 
However, the electromagnetic flux 
of energy at infinity through the event  horizon is positive almost 
everywhere with the exception of a very thin equatorial belt where 
it can be slightly negative (fig.\ref{NoPenrose}) . Thus, the pure 
electromagnetic Blandford-Znajek mechanism continues to operate. 
Moreover, the total flux of energy at infinity 
through the event horizon is positive and, in spite of the fact that 
the plasma magnetization is many orders of magnitude lower than that 
expected in typical astrophysical conditions, the BZ mechanism allows 
to extract the rotational energy of the black hole. 
 
One important feature of the initial phase which has been found in the 
previous simulations \cite{Koi03} and which persists in
the longer run is the lack of any noticeable plasma outflow from the black hole. 
The only exception is a thin sheath of radius $r \simeq 3$, 
most clearly seen in the top right panel of fig.\ref{final},) which is 
supplied with relativistically hot plasma 
($p/\rho \simeq 10^2$) via the reconnection process in the ergospheric 
current sheet.  
The outflow in the sheath is rather slow with poloidal speed relative to 
the local FIDO  never exceeding $v_p \simeq 0.7c$; it is most likely driven by 
the gas pressure.  This finding is in striking contrast with the results of 
MHD simulations of monopole magnetospheres of black holes which 
show the development of a powerful ultra-relativistic wind predominantly 
in the equatorial direction \cite{Kom04b}. Being taken together, these two results 
suggest that divergence of magnetic field lines is a necessary condition for 
generating strong ultra-relativistic plasma outflows from black holes.  
Moreover, given the fact that the Lorentz factor of the monopole wind reaches 
the value of $W=3$ in the equatorial direction only at the distance of 
around r=20 (or $20r_g$, where $r_g=GM/c^2$, if one prefers this dimensional 
form) and much further away in the polar  
direction, one would not expected to find an MHD-driven 
collimated ultra-relativistic outflow at distances smaller than several tens  
of $r_g$. Here there may be present only relativistic beams of particles 
accelerated by other mechanisms, e.g. electromagnetically. 
This makes the ongoing projects of studying the very bases (up to few tens 
of $r_g$) of astrophysical jets via short-wavelength VLBI observations  
particularly interesting \cite{Krich}. 

\begin{figure}
\includegraphics[width=80mm]{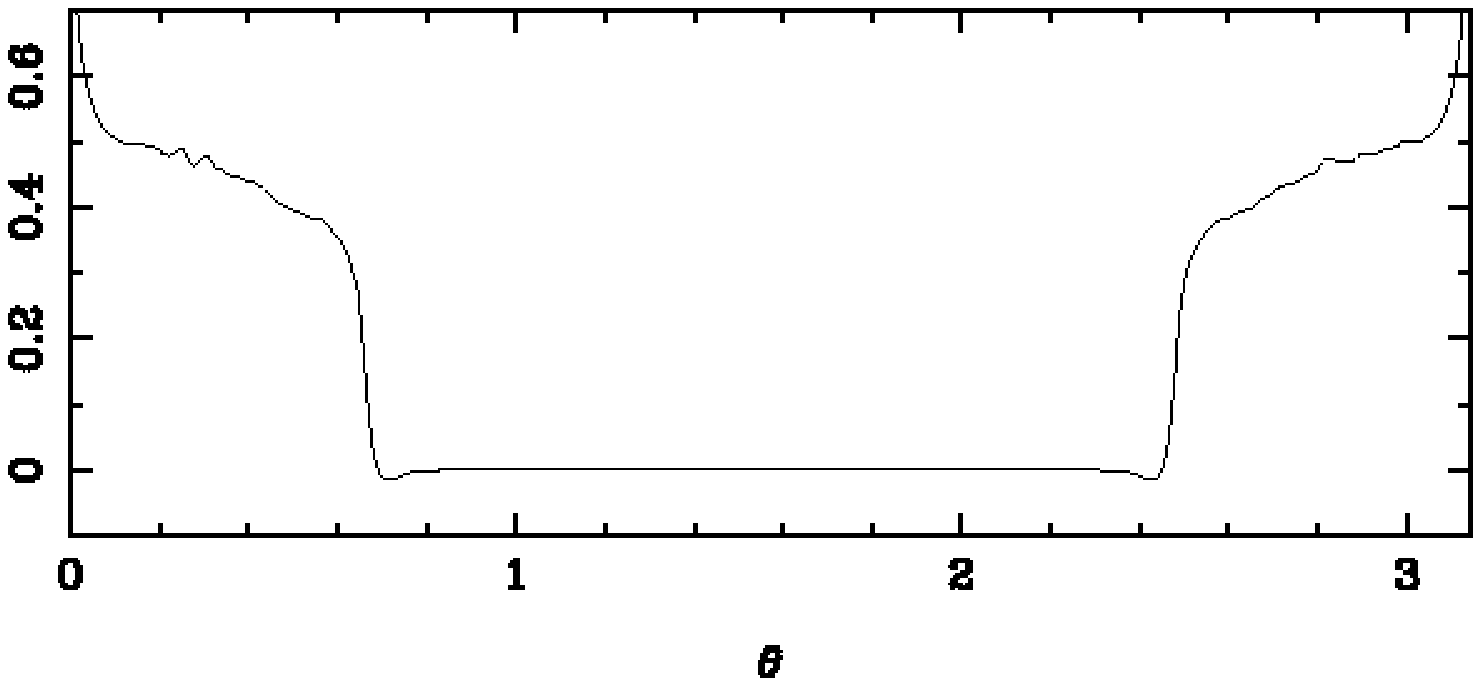}
\includegraphics[width=80mm]{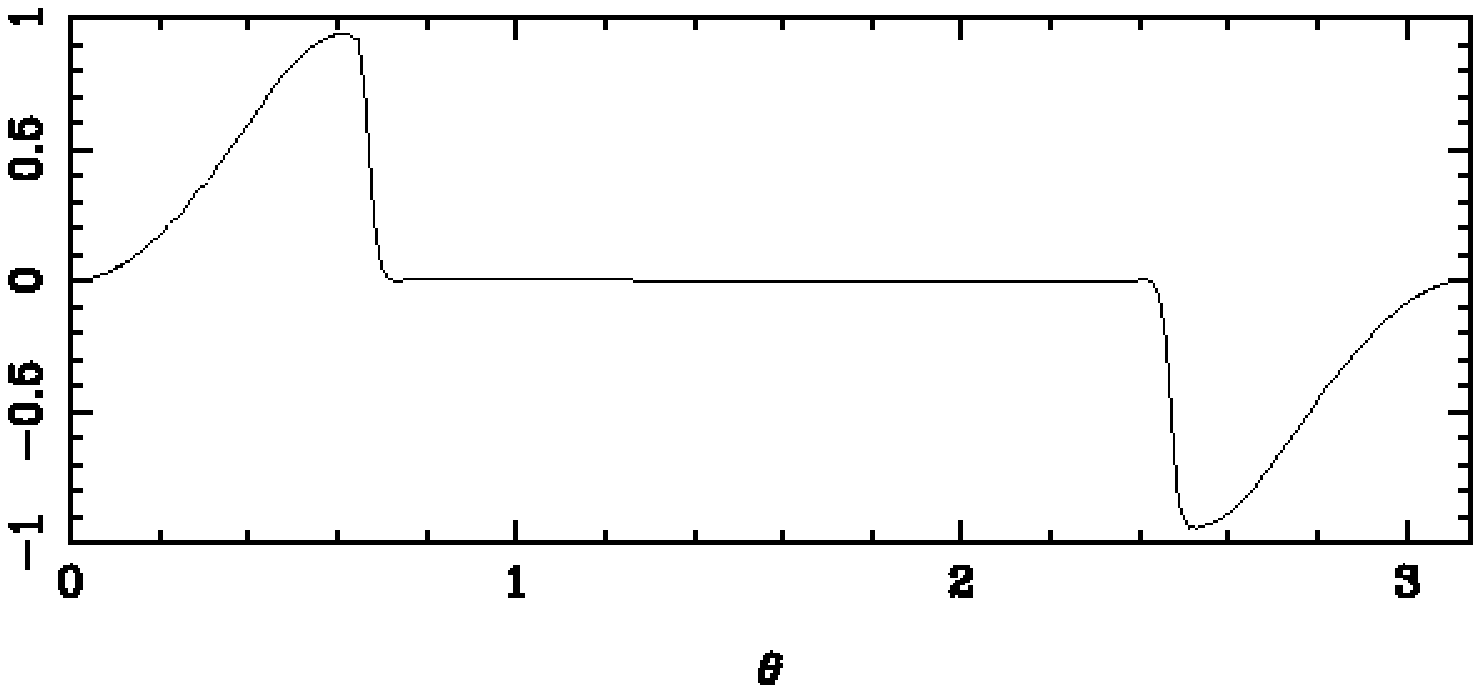}
\caption{ $\Omega/\Omega_h$ (top panel) and $H_\phi$ (bottom panel) at r=5 and t=60.}
\label{cuts}
\end{figure}

The bottom right panel of fig.\ref{final} shows the angular velocity of
magnetic field lines, $\Omega$, normalised to the angular  velocity 
of the black hole $\Omega_h=a/(r_+^2+a^2)$, where $r_+$ is the radius 
of the event horizon. In a steady state this parameter must be constant 
along magnetic field lines (e.g. Camenzind 1986) which is what is seen in 
the plot.  Thus, at $t=60$ our solution is very close to a steady 
state at least for $r < 20$. Another interesting result seen in this plot 
is that $\Omega \simeq 0.5 \Omega_h$, the value obtained in \cite{BZ} 
for a slowly rotating black hole with monopole magnetic field.       
Finally, there is a sharp transition between the ``rotating column''
of magnetic field lines attached to the black hole and the nonrotating
''soup'' of  magnetic field lines which fail to enter the ergosphere 
(see also fig.\ref{cuts}).   
This is most likely a discontinuity somewhat smeared by numerical diffusion 
and perhaps by the process of reconnection in the equatorial current 
sheet described above. 

The bottom left panel of fig.\ref{final} shows the distribution of 
$H_\phi$, the covariant azimuthal component of vector $\bH$ introduced
via 
$$
   H_i = \Fs_{ti} . 
$$  
In a steady-state 
$$
   \vcurl{H}=J,
$$
where $\bJ$ is the electric current density \cite{Kom04a}.
Thus,  $H_\phi$ at the point $(r,\theta)$ gives us the total electric 
current flowing through the loop $r=$const originated from this point. 
In force-free steady-state solutions $H_\phi$ is also constant along 
magnetic field lines \cite{BZ,Kom04a}.  This is exactly what is seen 
in figure~\ref{final},  thus confirming that at $t=60$ the solution  
is almost force-free and very close to a steady state for $r<20$.  
The distribution of $H_\phi$ also exhibits a discontinuity at the boundary 
of the ``rotating column'' indicating a thin current sheet of 
return current. 

The split-monopole structure of magnetic field  found in these 
simulations within the ergosphere is in conflict with the electrodynamic 
simulations where the field lines exhibit a sharp turning point 
in the equatorial plane \cite{Kom04a}. There seem to be only two
possible reasons for this difference. First of all, neither the inertia 
nor the pressure of plasma particles are accounted for in the 
electrodynamic model. At first 
glance, this does not seem to be important as even in the current 
MHD simulations both are small compared to the mass-energy density 
and pressure of magnetic field almost everywhere. But not exactly 
everywhere. In the equatorial current sheet the gas pressure 
dominates and plays a stabilizing role against otherwise quick 
reconnection of magnetic field lines. The second factor is the 
resistivity model. While the electrodynamic simulations \cite{Kom04a} 
utilize the anisotropic resistivity based on the inverse Compton scattering
of background photons, the only resistivity present in the MHD 
model is numerical. At this point we cannot exclude that explicit 
inclusion of physical resistivity will have a strong effect 
on the outcome of MHD simulations.  This matter will have to be addressed
in future studies.

\section{Summary and Conclusions}

We have carried out axisymmetric ideal MHD simulations of plasma 
flows in the case of a rotating black hole immersed into 
an initially uniform magnetic field described by the vacuum 
solution of Wald \shortcite{Wald}. Our main intention was 
to verify the results of previous simulations \cite{Koi02,Koi03} 
claiming a rather significant role for the so-called MHD Penrose 
process of extracting the rotational energy of black holes, at least 
in this particular case.  Our simulations show that this is true 
only for a short initial period during which a large fraction 
of magnetic field lines entering the black hole 
ergosphere exhibit a turning point in the equatorial plane. 
Eventually, those field lines are pulled into the black hole 
and within the ergosphere the magnetic field acquires 
the split-monopole structure. After this transient phase the 
regions of negative hydrodynamic energy at infinity are no longer present 
in the ergosphere and the the MHD Penrose process ceases to 
operate altogether. 

The rotational energy yet continues to 
be extracted via purely electromagnetic Blandford-Znajek process 
\cite{BZ,Kom04a}.  This energy extraction, however, is not followed by 
development 
of large-scale outflow from the black hole in sharp contrast with 
the results of previous MHD simulations for the monopole configuration
of magnetic field lines \cite{Kom04b}. The only exception is 
a thin current sheet located at the interface between the ''rotating
column'' of magnetic field lines attached to the black hole and 
the ''nonrotating soup'' of field lines failing to enter the 
black hole ergosphere. The relatively slow outflow in this 
''sheath'' is likely to be driven by gas pressure and is fed 
by hot plasma produced during  reconnection events in the 
equatorial current sheet.  Given the results of these and other 
recent MHD simulations \cite{Koi03,Kom04b} it seems impossible to generate 
via the standard MHD mechanism an outflow which becomes both 
ultrarelativistic and collimated already within the few first decades of 
the gravitational radius from a black hole. Should relativistic  
astrophysical jets continue to be seen at such small distances 
\cite{Krich}, one will have to look for  other explanations
of their origin.               

One of the main shortcoming of these simulations is the approximation 
of perfect conductivity. The only source of resistivity governing 
the process of magnetic reconnection in the developing current sheets
is purely numerical. Future numerical models will have to include physical 
resistivity.


\begin{thebibliography}{} 

\bibitem[\protect\citename{Beskin et al.}1991]{Bes}
  Beskin V.S., Istomin Y.N., Pariev V.I., 1991, Sov.Astron.,
  {\bf 36(6)}, 642.

\bibitem[\protect\citename{Beskin }2003]{Bes03}
  Beskin V.S., 2003, Phys.Uspekhi, {\bf 173}, 1247.  
                                                    
\bibitem[\protect\citename{Bicak \& Janis }1985]{BJ}
Bicak J. and Janis V., 1985, MNRAS, {\bf 212}, 899.        
 
\bibitem[\protect\citename{Blandford \& Znajek }1977]{BZ} 
 Blandford R.D. and R.L. Znajek R.L., 1977, MNRAS, {\bf 179}, 433.  

\bibitem[\protect\citename{Camenzind }1986]{Camen}
Camenzind, M., 1986, A\&A, {\bf 162}, 32.  

\bibitem[\protect\citename{Del Zanna et al.}2003]{Del-Zan} 
Del Zanna L., Bucciantini N., Londrillo P., 2003, 
A\&A, {\bf 400}, 397.

\bibitem[\protect\citename{De Villiers \& Hawley }2003]{Dev-Haw03} 
 De Villiers J.-P., Hawley J.F., 2003, ApJ, {\bf 589}, 458. 

\bibitem[\protect\citename{De Villiers et al. }2003]{Dev03} 
 De Villiers J.-P., Hawley J.F., Krolik J.H. 2003, ApJ, {\bf 599}, 1238.

\bibitem[\protect\citename{Evans \& Hawley }1988]{Eva-Haw} 
 Evans C.R., Hawley J.F., 1988, ApJ, {\bf 332}, 659. 

\bibitem[\protect\citename{Gammie et al.}2003]{Gam}
 Gammie C.F., McKinney J.C., T\'oth G., 2003, ApJ, {\bf 589}, 444. 

\bibitem[\protect\citename{Hirotani \& Okamoto }1998]{Hir-Oka}
 Hirotani K., Okamoto I., 1998, ApJ, {\bf 497}, 563.

\bibitem[\protect\citename{Koide et al.}1999]{Koi99}
 Koide S., Shibata K., and Kudoh T., 1999, Ap.J., {\bf 522}, 727. 

\bibitem[\protect\citename{Koide et al.}2002]{Koi02}
 Koide S., Shibata K., Kudoh T., Meier D.L., 2002, 
 Science, {\bf 295}, 1688.
                                                                                
\bibitem[\protect\citename{Koide }2003]{Koi03}
 Koide S., 2003, Phys.Rev.D, {\bf 67}, 104010.
 
\bibitem[\protect\citename{Koldoba et al.}2002]{Kol}
  Koldoba A.V., Kuznetsov O.A., Ustyugova G.V., 2002,
  MNRAS, {\bf 333}, 932.  

\bibitem[\protect\citename{Komissarov }1999]{Kom99} 
 Komissarov S.S., 1999, MNRAS, {\bf 303}, 343.  
                                                                                
\bibitem[\protect\citename{Komissarov }2001]{Kom01} 
 Komissarov S.S., 2001, MNRAS, {\bf 326}, L41  
                                                                                
\bibitem[\protect\citename{Komissarov }2002]{Kom02}
 Komissarov S.S., 2002, MNRAS, {\bf 336}, 759.

\bibitem[\protect\citename{Komissarov }2004a]{Kom04a} 
 Komissarov S.S., 2004a, MNRAS, {\bf 350}, 427.

\bibitem[\protect\citename{Komissarov }2004b]{Kom04b} 
 Komissarov S.S., 2004b, MNRAS, 350, 1431.

\bibitem[\protect\citename{Krichbaum et al. }2004]{Krich} 
 Krichbaum T.P. et al., 2004, astro-ph/0411487. 

\bibitem[\protect\citename{McKinney \& Gammie }2004]{McK} 
McKinney J.C., Gammie C.F., 2004, ApJ, {\bf 611}, 977. 

\bibitem[\protect\citename{Macdonald \& Thorne }1982]{MT82} 
 Macdonald D.A. and K.S.Thorne K.S., 1982, MNRAS, {\bf 198}, 345  
                                                                                
\bibitem[\protect\citename{Phinney }1982]{Phi}
 Phinney E.S., 1983, in Ferrari A. and Pacholczyk A.G. eds,
 {\it Astrophysical Jets}, Reidel, Dordrecht, p.201.

\bibitem[\protect\citename{Pons et al.}1998]{Pons} 
 Pons J.A., Font J.A., Ibanez J.M., Marti\'i J.M., 
 Miralles J.A., 1998, A \& A, {\bf 339}, 638 
                                                                                
\bibitem[\protect\citename{Priest \& Forbes}2000]{PF00}
Priest E. and Forbes T., ''Magnetic Reconnection'', Cambridge
University Press, Cambridge.  
                                                                                
\bibitem[\protect\citename{Punsly }2001]{Pun01}
 Punsly B., 2001, ``Black Hole Gravitohydromagnetics'', Springer-Verlag,
 Berlin.
 
\bibitem[\protect\citename{Punsly \& Coroniti} 1990a]{Pun-Cor} 
 Punsly B. and Coroniti F.V., 1990a, ApJ., {\bf 350}, 518. 

\bibitem[\protect\citename{Takahashi et al.}1990]{Tak} 
 Takahashi M., Niita S., Tatematsu Y., and Tomimatsu A., 
 1990, ApJ., {\bf 363}, 206.  
                                                                                
\bibitem[\protect\citename{Thorne et al.}1986]{TPM}
 Thorne K.S., Price R.H., and Macdonald D.A., 1986,
 ``The Membrane Paradigm'', Yale Univ.Press, New Haven.
                                                                                
\bibitem[\protect\citename{Wald }1974]{Wald}
 Wald R.M., 1974, Phys.Rev D, {\bf 10(6)}, 1680.
                                                                                
\end{thebibliography}
\end{document}